\documentclass[conference]{IEEEtran}
\IEEEoverridecommandlockouts
\usepackage{cite,amsmath,graphicx,booktabs,siunitx,xcolor}
\usepackage{gensymb}
\usepackage[hidelinks]{hyperref}
\usepackage{amssymb,amsfonts}
\usepackage{algorithmic}
\usepackage{textcomp}
\usepackage{graphicx}
\usepackage{svg}
\usepackage{orcidlink}

\newcommand{\NEligible}{1,326}

\newcommand{\NFbbPoints}{38,640}
\newcommand{\NTranSamples}{201,376}

\newcommand{\FzeroB}{1.3}
\newcommand{\FhalfB}{37.9}
\newcommand{\FfullB}{132.9}

\newcommand{\FRTpass}{326}

\begin{document}

% Working title. "Complete" is deliberately absent and may be inserted
% before "Commercial" only once the 1326-cell QA passes -- see
% TITLE_SELECTION.md and HOTFIX_FULL_LIBRARY_PRIOR_ART_MATRIX.md.
\title{Cryogenic Recharacterization of a Commercial 22\,nm FD-SOI
       Foundry Standard-Cell Library}

% ONE deliberate TODO remains in this manuscript, and it is this one.
% Names, order and affiliations are not recorded anywhere in the
% repository and are not guessed.  ISCAS review is NOT double-blind, so
% this block must be filled before submission.
\author{
\IEEEauthorblockN{
Mohamed~Gharib\,\orcidlink{0009-0008-3606-6200}\textsuperscript{1,2},
Rami~Rasheedi\,\orcidlink{0009-0009-8410-8011}\textsuperscript{1,2},
Salma~Abdelzaher\,\orcidlink{0009-0003-7242-5665}\textsuperscript{1,2}
Henry~Shi\,\orcidlink{0000-0001-6744-1941}\textsuperscript{2},\\
Mike~Hammer\,\orcidlink{0000-0003-0395-8225}\textsuperscript{2},
Llorenç Fanals-i-Batllori\textsuperscript{3},
Aldo Pena Perez\textsuperscript{3},
Antonino~Miceli\,\orcidlink{0000-0001-5994-5402}\textsuperscript{2},
and Inna~Partin-Vaisband\,\orcidlink{0000-0002-6399-6672}\textsuperscript{1}
}

\IEEEauthorblockA{
\textsuperscript{1}Department of Electrical and Computer Engineering,
University of Illinois Chicago, Chicago, IL, USA\\
\textsuperscript{2}Argonne National Laboratory, Lemont, IL, USA
\textsuperscript{3}SLAC National Accelerator Laboratory, Menlo Park, CA, USA
}
\thanks{This work was supported by U.S. Department of Energy
under Grants  DE-AC02-06CH11357 and DE-AC02-76SF00515.
}
\thanks{Corresponding author: Mohamed Gharib
(e-mail: mghari5@uic.edu).}
}

\maketitle

\begin{abstract}
\boldmath
Wiring and bandwidth bottlenecks between a quantum processor and
room-temperature electronics can be reduced by placing digital control at
the cryogenic stage, but commercial foundry standard-cell libraries are
not specified for operation at 4\,K. A cryogenic recharacterization flow
is therefore developed for a commercial 22\,nm FD-SOI foundry
standard-cell library using commercially available cryogenic
compact-device models, while retaining the original cell circuits, layouts, logic functions,
and implementation semantics and remaining compatible with the commercial
synthesis, place-and-route, and signoff methodology used at room temperature. Body-bias dependence is quantified over
\NFbbPoints{} matched slew/load points at zero, $\pm$0.9\,V, and
$\pm$1.8\,V body bias. Relative to full bias, zero-bias operation
increases the median rising and falling delays by $38.6\times$ and
$33.0\times$, while approximately 90\,\% of transitions exceed the
foundry-defined maximum-transition envelope. The 0.8\,V-to-0.4\,V level shifter is even more sensitive to body bias: 
its worst-case falling delay decreases from
30.6\,ns at zero bias to 0.56\,ns at $\pm$1.8\,V. For the outbound
0.4\,V-to-0.8\,V crossing, driving the 0.8\,V receiver without a level
shifter dissipates \SI{57.7}{\micro\watt} of static power per bit at full
bias, against 0.196\,nW when level-shifted. In post-route analysis of the
unchanged extracted physical database of a fabricated 32-bit ALU, the
register-to-register setup boundary increases from \FzeroB{}\,MHz at zero
bias to \FhalfB{}\,MHz at $\pm$0.9\,V and \FfullB{}\,MHz at $\pm$1.8\,V.
\end{abstract}

\begin{IEEEkeywords}
cryogenic CMOS, FD-SOI, standard-cell characterization, forward body bias,
level shifter, voltage domain, static timing analysis
\end{IEEEkeywords}

%=====================================================================
\section{Introduction}
%=====================================================================
Cryogenic classical logic is an important enabler for scalable quantum
processors and cryogenic sensor arrays, since control, decoding, and
readout placed near the cold-stage device can alleviate the wiring and
bandwidth bottlenecks to room-temperature
instrumentation~\cite{bardin2019,vandijk2020,sebastiano2025}. Compact
device models extending to 4\,K have been reported for several commercial
processes~\cite{beckers2019,chakraborty2021,xie2021}, and 22\,nm FD-SOI
devices have been characterized and modeled down to deep-cryogenic
temperatures~\cite{dutta2025}.

Device models alone, however, do not provide a digital implementation
flow. Synthesis, clock-tree construction, place-and-route, and signoff
operate on standard-cell libraries containing timing, transition, power,
and constraint data rather than directly on transistor compact models.
Measured 4\,K libraries have been demonstrated, but remain custom and
limited: a 40\,nm bulk library of nine cell types and seventeen drive
variants with resized and restacked cells~\cite{schriek2020}, and a
40\,nm bi-voltage library excluding odd-stage logic~\cite{lyu2024}.
Existing libraries have also been recharacterized for approximately
200 cells of an open-source 7\,nm kit at 10\,K~\cite{genssler2023} and
for production libraries down to 100\,K~\cite{saligram2024}. Body bias
has separately been established as an implementation variable in 22FDX
at room temperature~\cite{hoeppner2020} and evaluated for cryogenic
FD-SOI at the device and ring-oscillator levels~\cite{tdutta2026b}.

The approach considered here retains the commercial foundry cells without
cryogenic redesign and carries their sequential, clock-gating, test, and
multi-voltage semantics into a 4\,K commercial design flow. The complete
timing-eligible foundry library comprises \NEligible{} cells. FD-SOI is particularly suited
to this approach because the buried oxide separates the channel from the
back gate, avoiding the source--bulk junction-conduction constraint of
bulk triple-well technologies and permitting substantially larger forward
body bias~\cite{schriek2020}. This additional bias range can compensate
the cryogenic threshold-voltage increase~\cite{bohuslavskyi2018} without
requiring changes to the cell layouts.

Cryogenic standard-cell views are generated at 4\,K and a 0.4\,V core
supply under zero, $\pm$0.9\,V, and $\pm$1.8\,V body bias. The low core
supply is favored because it reduces switching energy and the associated
heat load on the limited 4\,K cooling budget, while forward body bias is
used to recover the performance otherwise lost to the cryogenic threshold
shift. The available I/O pads convert 1.8\,V external signaling to a
0.8\,V internal I/O domain, requiring explicit level translation between
that domain and the 0.4\,V logic core. 
The recharacterized libraries are
evaluated at both the voltage-domain interface and full-design levels using
a fabricated 32-bit 22\,nm FD-SOI ALU. The same extracted physical database
is retained across the cryogenic library conditions, enabling direct
comparison without changes to the implemented design.

%=====================================================================
%=====================================================================
\section{Characterization}
\begin{figure}[t]
\centering
\includegraphics[width=0.8\columnwidth]{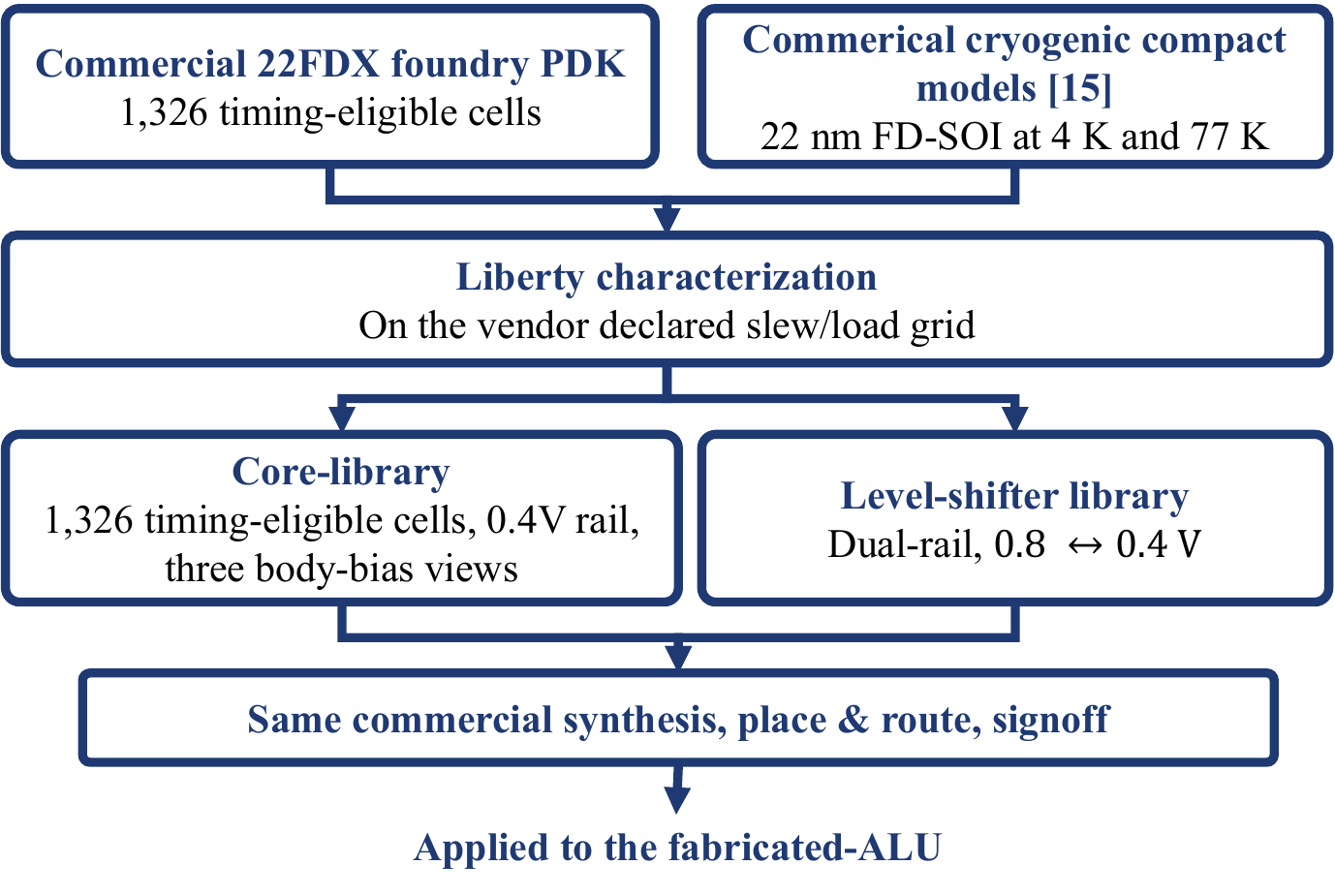}
\vspace{-10pt}
\caption{Characterization and implementation flow. The \NEligible{} timing-eligible cells are recharacterized at 4 K and 0.4 V, with body-bias dependence evaluated across three bias conditions. No
cryogenic cell redesign is performed and foundry semantics are
retained.}
\label{fig:flow}
\end{figure}

\begin{figure}[t]
\centering
\includegraphics[height=1.26in]{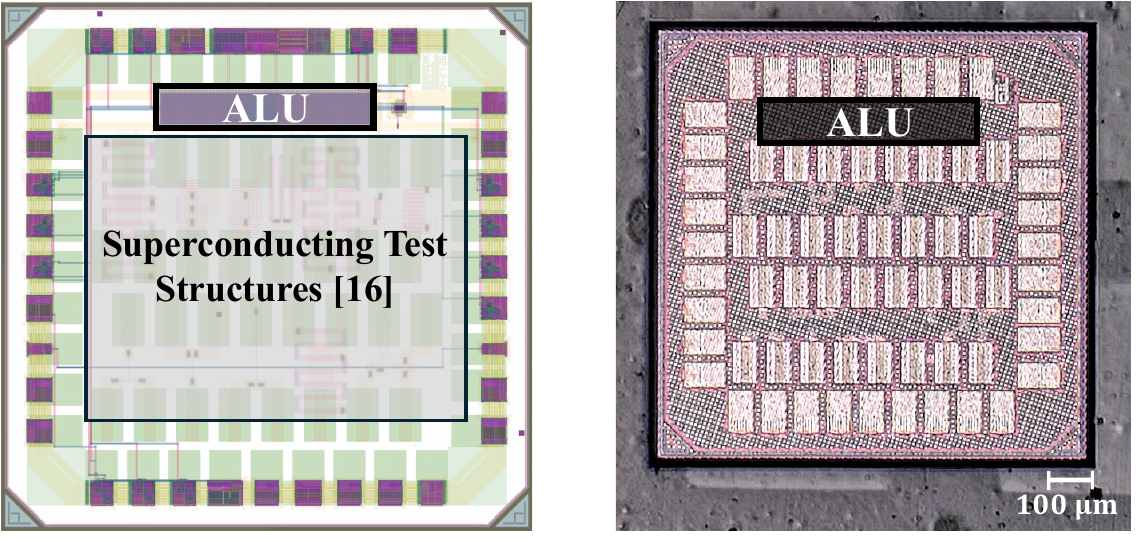}
\caption{Fabricated GF22FDX test chip. Left: physical layout. Right:
optical micrograph of the received die, 100\,\si{\micro\metre} scale bar.
The outlined 32-bit ALU occupies
399.97\,$\times$\,58.96\,\si{\micro\metre} of the
1000\,$\times$\,1000\,\si{\micro\metre} die; the remainder holds
superconducting test structures outside the scope of this
work~\cite{gharib2026}.}
\label{fig:die}
\end{figure}

%=====================================================================
Fig.~\ref{fig:flow} summarizes the cryogenic characterization and implementation flow, and Fig.~\ref{fig:die} shows the fabricated design. Standard cells from a commercial 22\,nm FD-SOI PDK are preserved without modifications to their topology, functionality, or layout. Cryogenic Liberty views are regenerated using foundry-extracted transistor netlists, independently supplied cryogenic compact-device models \cite{semiwise_cryopdk}, and a commercial characterization tool. The models are not foundry-qualified; therefore, the comparison validates the characterization flow rather than the underlying device models.
Timing, transition, and power tables are generated
on the vendor-defined slew/load grid and are consumed by the same
commercial synthesis, place-and-route, and signoff flow used at room
temperature.

Characterization is performed at 4\,K and $V_{DD}=0.4$\,V under three
back-gate conditions. Here, $V_{NW}$ and $V_{PW}$ denote the n-well
and p-well back-gate bias voltages, respectively. Zero bias and two levels
of forward body bias are considered. The full-bias campaign is performed over the complete
timing-eligible library, and body-bias trends are extracted at all three
bias conditions, so that each reported ratio is formed over a common cell
set.

The generated Liberty tables are independently checked against direct
transistor-level simulation. Representative cells are re-simulated from
the characterization decks with explicit waveform-based delay
measurements under identical device models and operating conditions. The
median delay discrepancy is approximately 0.065\,\%, with the 90th
percentile below 0.6\,\%, confirming consistency between the generated
tables and the underlying simulator. This comparison validates the
characterization flow, rather than the cryogenic device model itself.

%=====================================================================
\section{Body-Bias Dependence of Cell Timing}
\label{sec:cells}
%=====================================================================

Body-bias dependence is evaluated only at matched characterization points:
the same cell, pin, timing arc, input-slew index, and output-load index are
compared across bias conditions before population statistics are formed.
This procedure yields \NFbbPoints{} matched points available at all three
body-bias conditions and avoids comparisons between independently averaged
cell populations.

\begin{figure*}[t]
\centering
\includegraphics[width=0.9\textwidth]{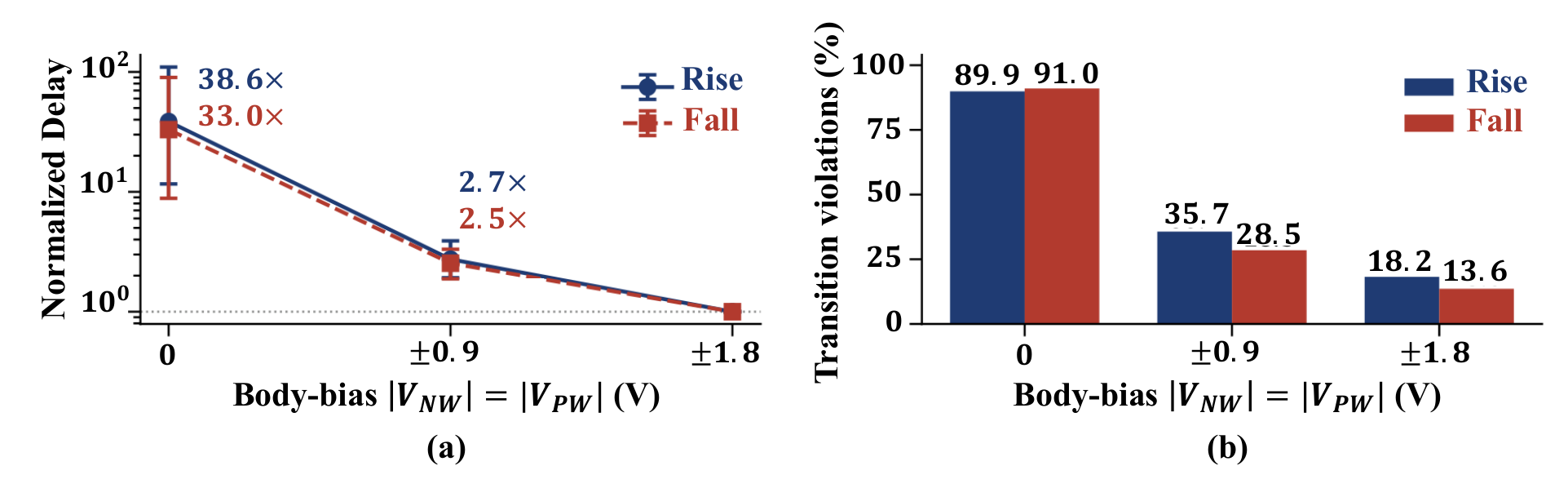}
\vspace{-10pt}
\caption{Cell timing dependence on forward body bias at 4\,K and 0.4\,V,
evaluated over \NFbbPoints{} matched slew/load points. (a)~Delay normalized to the $\pm$1.8\,V condition;
markers denote medians and bars indicate the 10th--90th percentile range.
(b)~Fraction of \NTranSamples{} characterized output transitions exceeding
the 0.7179\,ns per-pin \texttt{max\_transition} limit declared by the
vendor library.}
\label{fig:cells}
\end{figure*}

Fig.~\ref{fig:cells}(a) shows the delay distributions normalized to the
$\pm$1.8\,V condition. At zero bias, the median rising and falling delays
increase by $38.6\times$ and $33.0\times$, respectively. At $\pm$0.9\,V,
the corresponding penalties are reduced to $2.74\times$ and $2.54\times$.
The similar degradation observed on both transition polarities is
consistent with a library-wide back-gate effect rather than a delay
penalty confined to a single device polarity.

A wide cell-to-cell dependence remains within each bias condition. At zero
bias, the 10th-to-90th percentile delay range spans approximately one order
of magnitude, indicating that unbiased cryogenic timing cannot be
represented accurately by a single global derating factor.

Output transition exhibits an even stronger dependence on body bias and
directly determines whether operation remains within the characterized
slew range. The vendor library specifies a per-pin
\texttt{max\_transition} of 0.7179\,ns, corresponding to the upper bound of
the characterized slew grid, while the library-level
\texttt{default\_max\_transition} is 2\,ns. As shown in Fig.~\ref{fig:cells}(b), across
\NTranSamples{} characterized transitions, 89.9\,\% of rising and 91.0\,\% of falling
transitions exceed the per-pin limit at zero bias. These fractions decrease
to 35.7\,\% and 28.5\,\% at $\pm$0.9\,V, and to 18.2\,\% and 13.6\,\% at
$\pm$1.8\,V. Consistently, the median rising transition decreases from
8.07\,ns at zero bias to 0.095\,ns at full bias.

Such points would violate the library transition constraint and, when
propagated to a subsequent stage, can drive its input slew beyond the
characterized range. Forward body bias substantially reduces this
population, although some extreme grid points remain above the limit even
at $\pm$1.8\,V. The taped-out ALU, however, occupies only a subset of this
slew/load space; under full bias, none of its implemented paths violates
the maximum-transition constraint.

%=====================================================================
\section{The Voltage-Domain Interface}
%=====================================================================
Three voltage domains are present on the die. The pad ring operates at
1.8\,V, while the core-side interface of the foundry I/O cells operates
at 0.8\,V and the ALU core at 0.4\,V from a separate supply. The
0.8\,V domain is imposed by the specified operating range of the I/O
cells and cannot be collapsed onto the 0.4\,V core rail. Consequently,
every signal entering or leaving the ALU crosses the 0.8\,V/0.4\,V
boundary. Eight inbound and six outbound crossings are
implemented with the same level-shifter cell, with its supply connections
exchanged for the two directions.

\begin{figure*}[t]
\centering
\includegraphics[width=0.9\textwidth]{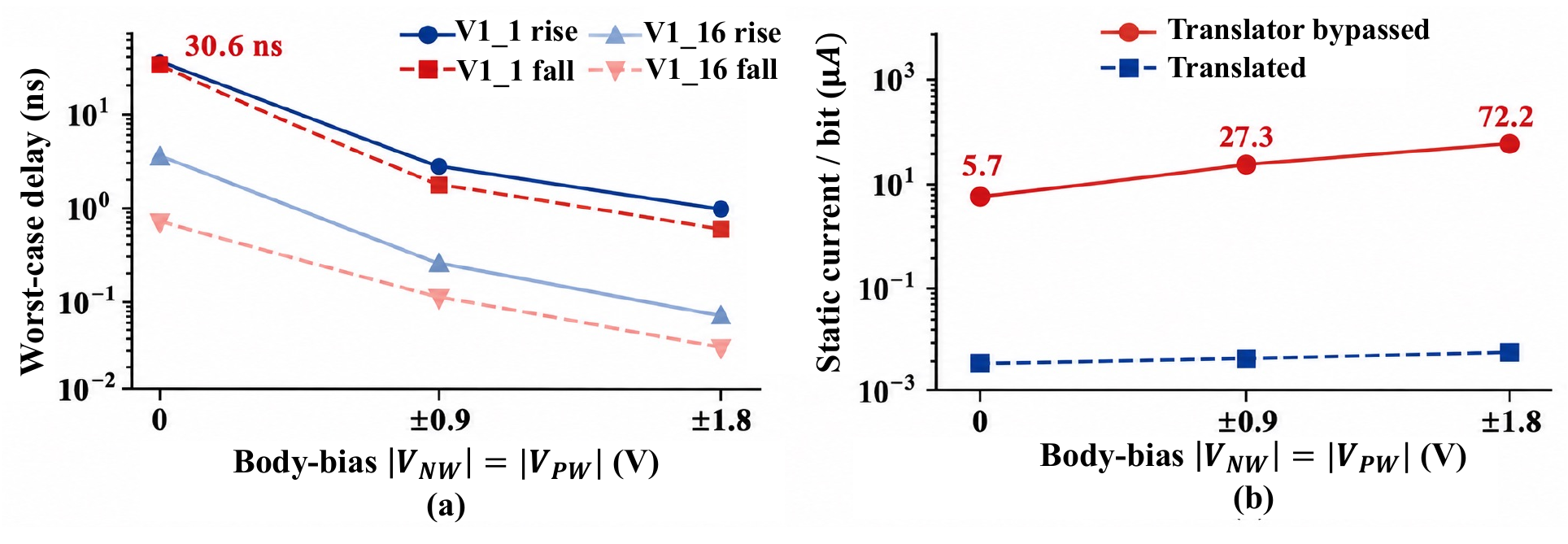}
\vspace{-10pt}
\caption{The voltage-domain translator at 4\,K. (a)~Worst-case delay of
the 0.8\,V\,$\rightarrow$\,0.4\,V translator against forward body bias,
for the smallest (\texttt{V1\_1}) and largest (\texttt{V1\_16}) drive.
(b)~Static current of the modeled outbound
0.4\,V\,$\rightarrow$\,0.8\,V crossing, translator bypassed and in place.
Bypassing under-drives the 0.8\,V receiver and draws
\SI{72.2}{\micro\ampere} per modeled path at $\pm$1.8\,V, against
sub-nanoampere when translated.}
\label{fig:ls}
\end{figure*}

Each crossing also includes a bypass path, with translated and untranslated
signals applied to a multiplexer controlled by a common select. This
structure permits the electrical cost of level translation to be evaluated
directly. Fig.~\ref{fig:ls}(a) shows the inbound translator delay versus
body bias. For the minimum-drive cell, the worst-case rising delay increases
from 0.803\,ns at $\pm$1.8\,V to 27.0\,ns at zero bias, while the falling
delay increases from 0.556 to 30.6\,ns. The resulting zero-bias penalties
of $33.6\times$ and $55.1\times$ exceed those observed for ordinary logic.
Increasing drive strength does not recover the unbiased response
proportionally; a $16\times$ drive increase improves the falling delay by
only $1.9\times$, indicating that output sizing alone does not remove the
cryogenic limitation.

The stronger requirement for translation is imposed by the outbound
0.4\,V-to-0.8\,V crossing. Fig.~\ref{fig:ls}(b) reports the static
boundary-circuit current when translation is bypassed and a 0.4\,V logic
high directly drives a representative 0.8\,V receiving stage. The complete
pad chain is not included, so the reported values correspond only to the
modeled domain boundary. Although the logic state is resolved, substantial
static conduction remains and increases with forward body bias. At
$\pm$1.8\,V, the bypassed path dissipates
\SI{57.7}{\micro\watt} per bit, compared with only 0.196\,nW for the
translated path. Across the six outbound crossings, this corresponds to
approximately \SI{346}{\micro\watt} for bypass operation versus 1.2\,nW
with translation.

The two directions therefore exhibit different constraints. On the inbound
0.8\,V-to-0.4\,V path, bypass operation is slightly faster and incurs little static
penalty, whereas on the outbound path it produces a substantial static-power
cost. Retaining the translated path is therefore the preferred operating
mode.

%=====================================================================
\section{Cryogenic Implementation-Level Performance and Energy}
%=====================================================================

The recharacterized libraries are applied to the routed 32-bit ALU using
the fabricated design's original netlist, parasitic extraction, and clock
tree. Placement, routing, extraction, crosstalk, on-chip variation,
common-path pessimism removal, and timing constraints are held fixed
across all cases; only the loaded cryogenic library is changed. Each
body-bias condition is then swept in clock frequency and analyzed
post-route.

Table~\ref{tab:chip} summarizes the resulting register-to-register setup
boundaries. At $\pm$1.8\,V, the design passes 132\,MHz with
$+0.053$\,ns setup slack. The corresponding
frequencies are 37\,MHz at $\pm$0.9\,V and
1.3\,MHz at zero bias. Accounting for the residual setup slack at
the highest passing point gives implied closure frequencies of
\FfullB{}, \FhalfB{}, and \FzeroB{}\,MHz, respectively. Thus, the same
physical implementation spans more than two orders of magnitude in setup
performance solely through the cryogenic operating condition.
\begin{table}[t]
\caption{Post-route analysis of the fabricated ALU at 4\,K and 0.4\,V
using one physical database and three characterized body-bias views.}
\label{tab:chip}
\centering
\footnotesize
\begin{tabular}{lccc}
\toprule
 & $0$\,V & $\pm0.9$\,V & $\pm1.8$\,V \\
\midrule
highest frequency passed      & 1.3\,MHz  & 37\,MHz   & 132\,MHz \\
setup slack at that pass (ns) & $+8.795$  & $+0.636$ & $+0.053$ \\
worst implied closure (MHz)   & \FzeroB{} & \FhalfB{} & \FfullB{} \\
hold slack (ns)               & $+33.578$ & $+1.363$ & $+0.319$ \\
transition violations (pins)  & 10,618    & 1,862     & 0 \\
worst transition (ns)         & 62.1      & 2.2       & $<0.7179$ \\
\bottomrule
\end{tabular}
\end{table}

The fabricated chip was signed off at 50\,MHz, the maximum operating
frequency specified for the 1.8\,V GPIO macro used in the tapeout, rather
than a limit of the ALU core. At 4\,K, the ALU exceeds this target only under full
bias: \FfullB{}\,MHz corresponds to $2.66\times$ the 50-MHz target,
whereas \FhalfB{}\,MHz is 24.2\,\% below it and zero-bias operation
reaches only 2.6\,\% of the target. The same extracted database evaluated
with its room-temperature signoff view at 0.65\,V and
25\,\celsius{} passes at \FRTpass{}\,MHz. The cryogenic full-bias boundary is therefore
approximately $2.45\times$ lower than the room-temperature boundary,
although the two points are not iso-voltage.

Hold timing remains positive at all three cryogenic bias conditions, with
margins of $+33.578$, $+1.363$, and $+0.319$\,ns from zero to full
bias. The full-bias margin is only 0.95\,\% of the zero-bias value,
showing that the bias condition used to recover setup performance must
also be included explicitly in hold signoff.

The cell-level transition behavior of Section~\ref{sec:cells} is also
observed on routed loads. At zero bias, 10,618 pins on 3,575
instances violate the declared maximum-transition limit, with a worst
transition of 62.1\,ns versus the 0.7179\,ns limit. At $\pm$0.9\,V,
1,862 pins on 1,069 instances remain in violation and the worst
transition is 2.2\,ns. No maximum-transition violations remain at
$\pm$1.8\,V. Full body bias therefore improves both setup performance
and operation within the characterized transition envelope.

\begin{figure}[t]
\centering
\includegraphics[width=0.9\columnwidth]{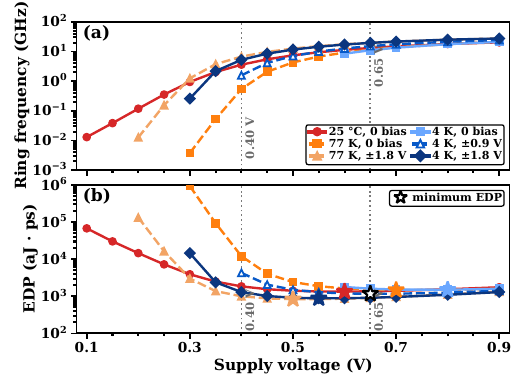}
\vspace{-5pt}
\caption{Seven-stage ring oscillator formed from the extracted commercial
foundry \texttt{INV\_1} cell at fanout one. (a)~Oscillation frequency
versus supply; each curve ends at the lowest tested supply sustaining
oscillation. (b)~Energy--delay product (EDP), minima starred. Forward
body bias reduces the minimum EDP by $1.52\times$ at 4\,K and
$1.60\times$ at 77\,K against the 25\,\celsius{} minimum. Dotted lines
mark the ALU's 0.4\,V cryogenic and 0.65\,V room-temperature supplies.
Core-supply power is used; the simulated body-bias contribution stays
below $3\times10^{-3}$\,\% of ring power. The cryogenic Liberty is
delivered at 0.4\,V only; every other supply shown is a direct
compact-model evaluation.}
\label{fig:edp}
\end{figure}

The ring oscillator in Fig.~\ref{fig:edp} separates supply-voltage and
body-bias effects. Without forward bias, the lowest tested supply that
sustains oscillation increases from 0.10\,V at 25\,\celsius{} to
0.30\,V at 77\,K and 0.60\,V at 4\,K. Full forward bias lowers these
cryogenic limits to 0.20\,V at 77\,K and 0.30\,V at 4\,K. The zero-bias ring and ALU results probe different
conditions: sustained oscillation requires closed-loop regeneration,
whereas STA evaluates externally driven paths, and the 0.4\,V zero-bias
ALU point also carries 10,618 maximum-transition violations. At matched
supply, the power--delay product varies by no more than 8\,\% across the
six temperature/bias conditions, while the 4\,K full-bias result remains
within 2.6\,\% of the 25\,\celsius{} value. Since
$\mathrm{PDP}=P_{\mathrm{avg}}t_{\mathrm{pd}}=E_{\mathrm{cyc}}/(2N)$ for an
$N$-stage ring, the frequency term cancels and PDP primarily reflects
switching energy.

The energy--delay product (EDP) therefore exposes the performance--energy
tradeoff more directly. Relative to the 25\,\celsius{} optimum, the
minimum EDP at 4\,K is $1.12\times$ higher without body bias but
$1.52\times$ lower at $\pm$1.8\,V. At 4\,K the optimum supply shifts
from 0.80\,V at zero bias to 0.55\,V at full bias. At the ALU's 0.4\,V operating point, the 4\,K
full-bias EDP is 1.298\,fJ\,ps compared with 1.848\,fJ\,ps at
25\,\celsius{}, corresponding to a $1.42\times$ reduction.

Direct transistor-level simulation of representative cells provides
an iso-voltage comparison with the foundry room-temperature model. At
0.4\,V, unbiased 4\,K delays are $7.1$--$71.6\times$ larger than at
25\,\celsius{}, whereas full forward bias reduces them to
$1.17$--$1.35\times$ below their room-temperature values. Static power
decreases from 1.8--1.9\,nW at 25\,\celsius{} to below 1\,pW at 4\,K.
At 100\,MHz, the corresponding per-cycle switching energies for these
cells differ by at most 2.1\,\% between the 4\,K full-bias and
25\,\celsius{} cases at the same 0.4\,V supply, indicating that the
dominant cryogenic change is delay rather than switching energy
\cite{vandijk2020,genssler2023,schriek2020}.

%=====================================================================
\section{Conclusion}
%=====================================================================
Using commercially available cryogenic compact-device models, a
commercial 22\,nm FD-SOI standard-cell library was recharacterized for
model-based 4\,K analysis at a 0.4\,V core supply without redesigning the
foundry cells. The resulting Liberty views retain the original cell
functions, layouts, and implementation semantics, and remain compatible
with the commercial digital implementation flow. Forward body bias
substantially improves cryogenic timing and output-transition compliance,
while explicit level translation avoids the static-power penalty of
driving the 0.8\,V domain directly from the 0.4\,V core. In post-route
analysis of the unchanged extracted physical database of a fabricated
32-bit ALU, the estimated register-to-register setup boundary increases
from \FzeroB{}\,MHz at zero bias to \FfullB{}\,MHz at $\pm$1.8\,V.
Library recharacterization, forward body bias, and explicit
voltage-domain translation can therefore support low-voltage cryogenic
digital implementation while preserving the foundry cell designs.
Cryogenic silicon measurements remain future work.

\clearpage
\bibliographystyle{IEEEtran}
\bibliography{references}

\end{document}